
\input phyzzx
\nopubblock

\overfullrule=0pt
\hsize=6 in
\vsize=8.5 in
\baselineskip=14pt
\line{\hfil August 1991}
\line{\hfil CCNY-HEP-91/11}
\vskip 0.5 in
\centerline{\bf COLLECTIVE FIELD REPRESENTATION OF NONRELATIVISTIC}
\centerline {\bf {FERMIONS IN (1+1) DIMENSIONS}} \footnote{*}{Talk presented at
the XXth
International Conference on Differential Geometric Methods in Theoretical
Physics at Baruch College,
CUNY, May 1991.}
\vskip 0.5 in \centerline{ DIMITRA KARABALI}\footnote{\dag}{e-mail
address:KARABAL@SCI.CCNY.CUNY.EDU}
\vskip 0.2 in
\centerline {\it Physics Department}
\centerline{\it  City College of the City University of New York}
\centerline {\it New York, NY 10031}
\vskip 0.7 in

\centerline{ABSTRACT}

 A collective field formalism for nonrelativistic fermions in
(1+1) dimensions is presented. The quantum mechanical fermionic problem is
bosonized and converted to a second quantized Schr\"odinger field theory. A
formulation in terms of current and density variables gives rise to the
collective field representation. Applications to the $D=1$ hermitian matrix
model
and the system of one-dimensional fermions in the presence of a weak
electromagnetic field are discussed.

\pageno=0
\vfil \eject

\hsize=6 in
\vsize=8.5 in
\baselineskip=14pt
\vskip 0.5 in
\centerline{\bf COLLECTIVE FIELD REPRESENTATION OF NONRELATIVISTIC}
\centerline {\bf {FERMIONS IN (1+1) DIMENSIONS}}
\vskip 0.5 in
\centerline{ DIMITRA KARABALI}
\centerline {\it Physics Department}
\centerline{\it  City College of the City University of New York}
\centerline {\it New York, NY 10031}
\vskip 0.5 in
\parindent=3pc
\midinsert
\narrower
\centerline{ABSTRACT}
\baselineskip=12pt

 A collective field formalism for nonrelativistic fermions in
(1+1) dimensions is presented. The quantum mechanical fermionic problem is
bosonized and converted to a second quantized Schr\"odinger field theory. A
formulation in terms of current and density variables gives rise to the
collective field representation. Applications to the $D=1$ hermitian matrix
model
and the system of one-dimensional fermions in the presence of a weak
electromagnetic field are discussed.
\endinsert
\parindent=2pc
\baselineskip=14pt
\vskip 1 cm
\noindent
1. {\bf Introduction}

The main motivation for studying one-dimensional nonrelativistic fermions came
from
the recent revival of interest in the matrix models in the framework of string
theory. In particular
the singlet sector of the one-dimensional $N \times N$ hermitian matrix model,
which provides the
triangulation for the $D=1$ string theory [1,9-14], is equivalent to a system
of
$N$ nonrelativistic fermions in an external potential [2].

In this talk we shall present a review of the collective field approach
to a quantum mechanical system of one-dimensional nonrelativistic fermions, in
terms of which a bosonic field theoretic description is derived.

Beyond the string theory context there is a variety of
condensed matter systems which are approximately described by low dimensional
nonrelativistic fermions  to which our formalism is also applicable.

After presenting the collective field formalism, we shall discuss its
application to the $D=1$ hermitian matrix model and the system of
one-dimensional fermions in the presence of a weak electromagnetic field.
Nontrivial soliton
solutions of the collective theory in the absence of an external potential are
discussed.

This talk is based on work done in collaboration with Bunji Sakita [3,4,5].

\vfill \eject
\noindent
2. {\bf{ Bosonization of Fermions}}

We start with a system of $N$ nonrelativistic fermions of mass $m$ described by
the Hamiltonian
$$ H  = {1 \over 2m} \sum_{
a=1}^{ N} ( -i \partial_{a} ~-~ {A}  ( {x}_{ a} ))^{ 2} ~+~ \sum_{ a}
V ( {x}_{ a }) ~+~ \sum_ { a \not= b } v ( {x}_{a} ~-~ {x}_{b} )  \eqno(1) $$
where $ {A} $ is
a gauge potential, $ V $ is a common potential and $v$ describes a two-body
interaction. The Schr$\ddot{\rm o}$dinger equation is given by
$$ H ~ \Psi ( {x}_{1} , {x}_{2} , \cdots , { x}_{N} ) ~=~E ~ \Psi
( {x}_{1} , {x}_{2} ,\cdots, { x}_{N} )
\eqno(2) $$
where $ \Psi ( x_{1} , {x}_{2} , \cdots , { x}_{N} )$ is a  totally
antisymmetric wave function.

The bosonization of the fermions is achieved by a singular
gauge transformation [3]
$$
\Psi ( {x}_{1} , {x}_{2} , \cdots , { x}_{N} ) ~=~ e^ { i
\Theta ( {x}_{1} , {x}_{2} , \cdots , { x}_{N} )} ~  \Phi
({x}_{1} , {x}_{2} , \cdots , { x}_{N} )
\eqno(3) $$
where $\Theta$ should be such that under an
interchange of a pair of variables:
$$
e^{i\Theta ( {x}_{1} , {x}_{2} , \cdots
,{x}_{N})}
\buildrel { {x}_{a}~ \leftrightarrow ~ {x}_{ b }} \over = -e^{ i\Theta (
{x}_{1} ,{x}_{2} ,
\cdots ,{ x}_{N} )} \eqno(4) $$ Then $\Phi$ is a totally symmetric wave
function. A realization of
$\Theta$ is given by  $$ \Theta ( {x}_{1} , {x}_{2} , \cdots , {x}_{N}) ~=~
\sum_{ a> b }~Im~ ln (
x_{a} ~-~ x_{ b } - i \epsilon ) \eqno(5) $$

Since (3) is a (singular) gauge transformation, the Hamiltonian for $\Phi$ is
obtained from (1) by adding the following (singular) gauge term to $A ({x}_{a}
) $
$$
a( x_{ a} ) ~=~-   \pi ( \sum_{ b > a} \delta ( x_{ a} - x_{b} )~-~
\sum_{ b < a} \delta ( x_{a }- x_{ b} ))
\eqno(6) $$

The bosonic Hamiltonian is now given by [3]
$$ \eqalignno{
H & = {1 \over 2m} \sum_{a=1}^{N} ( -i \partial_{ a} ~-~ A  ( x_{a })) ^{ 2}
{}~+~ \sum_{a}
V ( x_{ a}) ~+~ \sum_{ a \not= b } v ( x_{a} ~-~ x_{ b} )  \cr &~+~ {\pi^{2}
\over m} [ \sum_ { a> b
> c } \delta ( x_{ a} ~-~ x_{ b} )~ \delta ( x_{a} ~-~ x_{c} ) ~+~ \sum_{ a > b
}
\delta^{2} ( x_{a} ~-~ x_{ b} ) ] &(7) \cr} $$

The bosonization procedure gave rise to a three-body and a singular
two-body interaction.

\noindent
3. {\bf{ Quantized Schr\"odinger Bose Field Formalism}}

It is useful to express this many body system by using the second quantized
formalism. As usual we introduce a quantum field operator $\psi (x) $ and its
conjugate $ \psi^{\dag} ( x ) $ , which satisfy the Bose commutation
relation at equal times
$$ \eqalignno {
& [ \psi ( x ) , ~\psi ^{\dag} ( y ) ] ~=~ \delta ( x ~-~ y ) \cr
& [ \psi(x), ~\psi(y)] = [\psi^{\dag} (x), ~\psi^{\dag} (y)] =0 &(8) \cr} $$
We construct a symmetric basis vector
$$
| x_{ 1} , x_{ 2} , \cdots , x_{ N} > ~=~ {1 \over {(N ! )^{1/2}}}
\psi^{\dag}( x_{ 1}  ) \psi^{\dag} ( x_{ 2} ) \cdots \psi^{\dag} ( x_{ N} ) |0>
\eqno(9) $$
Since
$$ \eqalignno {
& \int dx dy \psi^{\dag} ( x ) \psi^{\dag} ( y ) V(x, y) \psi ( y )
\psi ( x ) | x_{ 1} , x_{ 2 }, \cdots , x_{ N} > ~=~\sum _ { a \not= b }
V(x_{a}, x_{b})  | x_{ 1 }, x_{ 2} , \cdots , x_{ N} > \cr
&  \int dx dy dz \psi^{\dag} ( x ) \psi^{\dag} ( y ) \psi^{\dag} (z) V(x, y,z)
\psi(z) \psi ( y ) \psi ( x ) | x_{ 1} , x_{ 2 }, \cdots , x_{ N} > = \cr
& \sum _{a \not= b \not= c} V(x_{a}, x_{b}, x_{c})  | x_{ 1 }, x_{ 2} ,
\cdots , x_{ N} >
&(10) \cr} $$
we find that the second quantized expression of eq.(7) is given by
$$ \eqalignno {
H & ~=~ \int dx  [ {1 \over 2m} ( D \psi (x) ) ^{\dag} ( D \psi (x)) ~+~
\psi ^{\dag}( x ) \psi ( x  ) V (x) ~+~{ \pi^{2}  \over 6m} (\psi ^{\dag } ( x
) \psi(x))^{3}  \cr
& -{\pi^{ 2} \over 6m} \delta^{2} (0) \psi^{\dag } ( x ) \psi ( x  ) ]
+\int dx dy
\psi^{\dag}( x ) \psi^{\dag} ( y  ) \psi ( y ) \psi ( x  ) v ( x ~-~y ) &(11)
\cr}
$$
where $D $ is a covariant derivative $ D ~=~ {\partial \over \partial x } ~-~i
{}~A(x) $.

An analogous procedure of bosonization and second quantization can be applied
to
a (2+1) dimensional system of nonrelativistic fermions. In this case one
obtains
a Hamiltonian for the bosonic field $\psi$ with an additional coupling to a
Chern-Simons gauge field [3].

\vfill \eject
\noindent
4. { \bf {Collective Field
Formalism}}

The Hamiltonian $H$ in eq.(11) can be
written as a functional of the charge density $\hat{\rho} (x) ~=~
\psi^{\dag}(x)
\psi (x) $ and the current density $ \hat{j} (x) ~=~ i \psi^{\dag}(x) \partial
 \psi (x) $
$$ \eqalignno{
& H[ \hat{j}  , \hat{\rho} ] = \int dx \bigl[ {1 \over 2m} \bigl( \partial
\hat{\rho} (x) + i \hat{j} (x) + i \hat{\rho} (x) A(x) \bigr)~ {1 \over
\hat{\rho}(x)} \bigl( -i \hat{j} (x) - i \hat{\rho} (x) A(x) \bigr) +
\hat{\rho}
(x) V(x) \cr &  + {\pi ^{2} \over 6m} \hat{\rho} ^{3} (x) - {\pi ^{2} \over 6m}
\delta ^{2} (0)
\hat{\rho}(x) \bigr] + \int dx dy \bigl( \hat{\rho} (x) \hat{\rho} (y) - \delta
(x-y) \hat{\rho} (x)
\bigr) v(x-y) &(12) \cr} $$
The idea of using the density of currents to describe the Schr$
\ddot{\rm o}$dinger wave field theory is due to Dashen and Sharp [6]. Our
definition of the current density is slightly different from theirs for later
convenience, see eq.(19). The Schr$\ddot{\rm o}$dinger equation is given by
$$ H ~|\Phi~> ~=~ E ~| \Phi ~> \eqno(13) $$
where $ | \Phi > $ is an $N$
particle Bose state expressed in terms of a symmetric wave function $ \Phi (
x_{
1} , x_{ 2} ,\cdots , x_{ N} ) $ as   $$ |\Phi>~=~ \int \cdots \int dx_{ 1}
dx_{
2} \cdots d x_{ N} ~ \Phi ~ ( x_{ 1} , x_{ 2} , \cdots , x_{ N} ) ~ | x_{ 1} ,
x_{ 2} , \cdots , x_{ N} > \eqno(14) $$
where $ | x_{ 1} , x_{ 2} , \cdots , x_{ N} >  $ is the basis
vector defined by eq. (9). Since $\Phi$ is a symmetric function, we may regard
it as a functional of the density variable, namely
$$
\Phi ( x_{ 1} , x_{ 2} , \cdots , x_{ N} ) ~=~ \Phi ~ [ ~\rho ~] \vert  _{ \rho
(x) ~=~ \sum_{ a} \delta ( x ~-~ x_{ a })  }
\eqno(15) $$
Thus,
$$ \eqalignno {
 |\Phi> & = \int \cdots \int dx_{ 1} dx_{ 2} \cdots d x_{ N} ~
\Phi [ \hat{\rho} ] ~ | x_{ 1} , x_{ 2} , \cdots , x_{ N} > \cr
& = \Phi [  \hat{\rho} ] {1 \over { ( N ! )^{1/2 }}} \bigl( \int dx
\psi^{\dag} (x) \bigr) ^ {N} |0> &(16) \cr} $$
where we used
$$
\hat{\rho} (x) ~ | x_{ 1} , x_{ 2} , \cdots , x_{ N} >~=~ \sum_{a} \delta ( x
{}~-~ x_{ a} ) ~ | x_{ 1}
, x_{ 2} , \cdots , x_{ N} > \eqno(17) $$

Next we use the following commutation relations of charge and current density
in
order to obtain the collective field representation of the Schr$\ddot
{\rm o}$dinger equation:
$$[ \hat{j} (x) , \hat{\rho} ( y ) ] ~=~ i \hat{\rho}(x) \partial_{ x} \delta
( x ~-~ y ) ~  ~,~~~~~~~~
[ ~\hat{\rho} (x) , ~\hat{\rho} ( y) ] ~=~ 0
\eqno(18) $$
Since
$$
\hat{j} (x) {1 \over { ( N ! ) ^{1/2}}} \bigl( \int dx \psi^{\dag} (x)
\bigr) ^{N} |0> ~= ~0
\eqno(19) $$
we obtain
$$ \eqalignno {
\hat{j} ( x ) | \Phi > ~& =~[ \hat{j} (x) , \Phi [ \hat{\rho}] ] {1 \over { ( N
! ) ^{1/2}} } \bigl( \int dx \psi^{\dag} (x) \bigr) ^ {N} |0> \cr ~& =~\bigl( ~
i \hat{\rho} ( x )
\partial_{ x } {\delta \over { \delta \hat{\rho}(x) }} \Phi [ \hat{\rho}]
\bigr) {1 \over { ( N
! ) ^{1/2}} } \bigl( \int dx \psi^{\dag} (x) \bigr) ^ {N} |0>
&(20) \cr} $$
Therefore the collective field representation of the Schr$\ddot {\rm o}$dinger
equation is given by
$$
H [ ~i \rho \partial_{ x} {\delta \over { \delta \rho }} , ~ \rho ] ~\Phi~[ ~
\rho ~] ~=~ E ~\Phi [ ~\rho ~]
\eqno(21) $$
This is not however the final form of the Schr$\ddot {\rm o}$dinger equation.
The transition to
the density variables $\rho (x)$ introduces a nontrivial Jacobian factor. The
inner product of the states in the functional space of $ \rho (x) $ is
$$
<~\Phi^{'} ~|~ \Phi~> = \int {\cal D} \rho ~J [ \rho ] \Phi ^{' *} [ \rho ]
\Phi [ ~\rho ~]
\eqno(22) $$
where $J[\rho ] $ is given by [3]
$$
J[ \rho ] = ~\int \cdots \int dx_{ 1} dx_{ 2} \cdots dx_{ N} \prod _{x} \delta
( \rho
(x) ~-~ \sum _{a} \delta (x ~-~ x_{ a} )) \eqno(23) $$
The Jacobian $J$ can be calculated order by order in $1/N$ expansion [3,4,7].
We
find that
$$
J [ \rho ] ~=~  \delta (
\int dx \rho (x) ~-~ N) ~ j [\rho]  $$
where
$$j[\rho]=exp \Bigl[- \int dx \rho (x) ln \rho (x) -{\delta (0) \over 2} \int
dx ln \rho
(x)-{\delta ^{2} (0) \over 12} \int {dx \over \rho (x)} + O(N^{-2}) \Bigr]
\eqno(24)
$$
 In the
functional space of $\rho (x)$ with measure ${\cal {D}} \rho (x)=\prod_{x}
d\rho (x) ~ \delta( \int dx \rho (x) -N)$, the wavefunction is defined through
the similarity transformation  $$
j^{1/2}\Phi [ \rho ] =\Psi [ \rho ] \eqno(25) $$
and the corresponding Hamiltonian is given by
$$
 j^{1/2}~ H[~ -i\rho
\partial {\delta \over \delta \rho}, \rho]~ j^{-1/2}= H [ \rho \partial \pi -{i
\over 2} \partial
\rho +\delta (0) {{i \partial \rho} \over 4 \rho}- \delta ^{2} (0) {{i \partial
\rho} \over 12 \rho
^{2}}+ \cdots, \rho ]
 \eqno(26) $$
where $ \pi (x) ~=~ - i {\delta \over { \delta \rho (x) }} $.

The second quantized form of the Hamiltonian can now be expressed in terms of
the canonical fields
$\rho (x)$ and $\pi (x)$ as
$$\eqalignno{ H = & \int dx \Bigl[{1 \over 2m} \rho (x) \Bigl(\partial \pi (x)
+i{{\delta (0)} \over 4}{{\partial \rho (x)} \over {\rho^{2}
(x)}}-i{{\delta^{2} (0)} \over 12} {\partial \rho(x) \over {\rho ^{3}
(x)}} + \cdots -A(x) \Bigr) ^{2}  \cr
& +  {1 \over 8m} {{(\partial
\rho (x))^ {2}} \over \rho (x)}  + {\pi ^{2} \over 6m} \rho ^{3} (x) + V(x)
\rho
(x) +{1 \over 4m} \delta^{''} (0) -{\pi ^{2} \over 6m} \delta ^{2} (0) \rho
(x)  \Bigr] \cr
& + \int dx dy \rho (x) \rho (y) v(x,y) - Nv(0) &(27) \cr} $$
where the density constraint $\int dx
\rho (x) =N$ was used.   This expression for the Hamiltonian
agrees, up to the subleading in $N$ singular terms introduced in the
calculation of the Jacobian, eq.(24), with the expression  derived by applying
the collective field approach of [8] on the bosonized version of the many-body
Hamiltonian (7).

Equation (27) provides now a bosonic field Hamiltonian amenable to a
semiclassical treatment.

\vskip 1 cm
\noindent
{\bf{5. Applications}}

\noindent
{\it {5.1. $D=1$ Matrix Model}}

The relation between one-dimensional nonrelativistic fermions and the $D=1$
hermitian matrix model has been established in the classic paper of
Br$\acute{\rm e}$zin,
Itzykson, Parisi and Zuber [2] whose analysis we briefly outline.

 The dynamics of the
$D=1$ hermitian matrix model is determined by the Lagrangian $$ L=tr \bigl[ {1
\over 2} \dot{M} ^{2} - V(M) \bigr] \eqno(28) $$ where $M$ is an $N \times N$
hermitian matrix and $V(M)$ is a polynomial in $M$. In particular we consider
functions $V(x)$ which scale with $N$ like $V(x) = NV(x/ \sqrt{N})$. One
quantizes the system in ``cartesian coordinates" defined by $M_{i} =
tr(Mt_{i}),~i=1, \cdots , N^{2}$, where $t_{i}$ is a hermitian basis of the
fundamental representation of $U(N)$ Lie algebra.

The Lagrangian (28) is invariant
under a time independent $U(N)$ transformation $M \rightarrow UMU ^{\dag}$,
hence it is natural to consider an analogue of the polar coordinate basis
$$M=UXU^{\dag},~~~~~~~X_{ab}=x_{a}\delta _{ab} \eqno(29)$$
The integration measure is given by
$$ \prod_{i} dM_{i} = \Delta^{2} (x) \Bigl(\prod_{a} dx_{a}\Bigr) dU
\eqno(30)$$
where $\Delta (x) \equiv \prod_{a>b}(x_{a}-x_{b})$ is the Vandermond
determinant.
We shall restrict our
attention to the singlet sector; the wave function
$\Omega$ is a symmetric function of the eigenvalues $x_a, ~a=1, ... , N $. The
Jacobian factor
$\Delta ^{2} (x)$, eq.(30), which appears in changing variables from the
``cartesian coordinates" to the ``polar coordinates" can be absorbed by
appropriately redefining the wavefunction and the Hamiltonian. This is done by
redefining an  antisymmetric wavefunction $$ \Psi (x_{1}, \cdots, x_{N}) =
\Delta
(x) \Omega (x_{1}, \cdots, x_{N}) \eqno(31) $$ The corresponding Schr$\ddot{\rm
{o}}$dinger equation
is  $$ \sum _{a} \bigl[ -{1 \over 2} {\partial ^{2} \over \partial x_{a} ^{2}}
+ V(x_{a}) \bigr]
\Psi (x_{1}, \cdots, x_{N}) = E \Psi (x_{1}, \cdots, x_{N}) \eqno(32)$$
Therefore the singlet sector
of the original matrix model can be equivalently described by a set of $N$
decoupled
nonrelativistic fermions in the external potential $V(x_{a})$ [2] .

According
to the collective field formalism developed earlier the field theoretic
Hamiltonian appropriate for describing the singlet sector of the $N$ hermitian
$D=1$ matrix model is given by
$$\eqalignno{ H  & = \int dx \Bigl[{1 \over 2} \partial \pi (x) \rho (x)
\partial
\pi (x)  +  {1 \over 8} {{(\partial \rho (x))^ {2}} \over \rho (x)}
+ {\pi ^{2} \over 6} \rho ^{3} (x) + V(x) \rho (x) \cr &
+{1 \over 4} \delta^{''} (0)
-{\pi ^{2} \over 6} \delta ^{2} (0) \rho (x)
+{{i \delta (0)} \over 8}
\{\partial \pi (x), ~{\partial \rho (x) \over \rho (x)}\}
-\delta^{2} (0) {(\partial \rho(x))^{2} \over 32\rho^{3} (x)} \cr &
-{i\delta^{2} (0) \over 24} \{\partial\pi (x),~{\partial \rho (x) \over
\rho^{2} (x)}\}
+\delta^{3}(0){(\partial \rho (x))^{2} \over 48\rho^{4}(x)}
+ \cdots \Bigr] -e(\int dx \rho (x)-N) &(33) \cr} $$
where $e$ is a Lagrange multiplier for the density constraint.

Of particular interest is the double scaling limit of the $D=1$ hermitian
matrix model. This provides a definition of two-dimensional gravity coupled to
a scalar field [1]. In the fermionic description the double scaling limit of
the
theory is taken by keeping the difference between the value of the potential at
the local maximum and the fermi energy fixed as $N \rightarrow \infty $, $V (
x_0 ) - e_F
\rightarrow \mu $ [9],[10].
Expanding the potential $V$ around the local maximum
$x_0$ and then taking the
large $N$ limit, one concludes that the double scaling limit of the $D=1$
hermitian matrix model is equivalent to a system of fermions in an inverted
harmonic oscillator potential [11],[12].
According to the previous discussion the corresponding bosonized
Schr\"odinger wave field Hamiltonian for this system is (up to a constant)$$
H= \int dx ~ \Bigl[{ 1 \over 2 } \partial {\psi ^{\dag}} \partial \psi
+ {{\pi ^2}\over{6}} {({\psi ^{\dag}} \psi
) } ^3 +( \mu- {{ x ^2}\over 2}
) {\psi ^{\dag}} \psi \Bigr] \eqno(34)$$
The corresponding collective field representation of the Hamiltonian is given
by $$H = \int dx \bigl[ {1 \over 2}  \partial \pi (x)  \rho(x)  \partial \pi
(x)  + {1 \over 8} {({\partial \rho (x)) ^{2}} \over {\rho (x)}}+{\pi ^{2}
\over
6} \rho^{3} (x) + (\mu -{x^{2} \over 2}) \rho(x) + \cdots \bigr]
\eqno(35) $$
where $\cdots$ contains the singular terms (see eq. 33).
We notice that the second term in (35),whose origin lies on the term $\int \rho
ln \rho$ of the Jacobian (24), is absent from the collective field Hamiltonian
of
[8] used by Das and Jevicki [13] for the discussion of the double scaling
limit.

Given the Hamiltonians (34), (35) one can apply a semiclassical analysis. Of
particular interest are nontrivial classical solutions and their interpretation
in the $D=1$ string theory framework.\footnote{*}{Classical solutions
of the cubic collective theory (without the $(\partial \rho) ^{2} / \rho$) have
been studied in refs.[14-16].}

The classical equations of motion corresponding to eq.(35) are
$$ \eqalignno{\dot {\rho} & = -\partial (\rho \partial \pi) \cr
-\dot{\pi} & = {1 \over 2} (\partial \pi) ^{2} +{1 \over 8} \Bigl[{{(\partial
\rho) ^{2}} \over
{\rho ^{2}}} -2 {{\partial ^{2} \rho} \over \rho} \Bigr] + {\pi ^{2} \over 2}
\rho ^{2} +\mu -{x^{2}
\over 2} &(36) \cr}$$
The equivalence of this set of equations of motion and the one derived from
(34)
is straightforward given that the classical Schr$\ddot{\rm o}$dinger wave field
and the collective field are related to each other by
$$ \psi (x) = \rho ^{1/2} (x) e^{i \pi (x)} \eqno(37) $$

The presence of the inverted harmonic oscillator potential complicates the
search for explicit
analytic solutions of (36). It was shown though in [16] that the harmonic
oscillator potential can be induced through a reparametrization. It is
interesting then to study solutions of eq.(36) without the harmonic oscillator
potential term [5,17]. In this case we find a nontrivial static solution
of the form $$\rho (x)=\rho _{0} \Bigl[1-{{36 \rho _{0}} \over {\alpha
^{-1}e^{2
\pi \rho _{0}x}+24 \rho _{0} + 36 \rho ^{2}_{0} \alpha e^{-2 \pi \rho _{0}x}}}
\Bigr] \eqno(38)$$ where $\rho _{0}={p_{F} \over \pi}$. Because of translation
invariance there is a free parameter $\alpha$, which controls the position of
the
minimum of the configuration. The derivative term $(\partial \rho)^{2}/ \rho$
is
very crucial for the existence of such a solution.

After some straightforward algebra we find that the energy of this static
soliton configuration is
$$E=H[\rho ]-H[ \rho _{0}]=p_{F}^{2} {\sqrt{3} \over {2 \pi}} log \Bigl({1
\over {2-\sqrt{3}}}\Bigr) \eqno(39)$$

The time-dependent solution corresponding to the moving soliton is of the form
[5] $$\rho (x,t)=\rho _{0} -{36 c \over {(\pi ^{4}\rho _{0} ^{2}+3 \pi ^{2}
v^{2})c^{-1}\alpha ^{-1}e^{-2 \sqrt{c}(x+vt)}+24 \pi ^{2} \rho _{0} + 36 c
\alpha
e^{2 \sqrt{c}(x+vt)}}}  \eqno(40)$$
where $c=\pi ^{2} \rho
^{2} _{0}-v^{2}$ and $|v| \le \pi ^{2} \rho
^{2} _{0}$. The corresponding energy is
$$E=H[\rho]-H[\rho _{0}]=(p_{F}^{2}-v^{2}) {\sqrt{3} \over {2 \pi}} log
{{\sqrt{p_{F} ^{2}+3v^{2}}}
\over {2p_{F}-\sqrt{3(p_{F}^{2}-v^{2})}}} \eqno(41)$$

These soliton solutions, although different from the ones discussed by
Jevicki in ref.[16], display similar features.
 \vskip 0.7 cm
 \noindent
{\it{5.2.
One-Dimensional Fermions in External Electromagnetic Field}}

We start with a system of $N$ nonrelativistic fermions in the presence of a
weak electromagnetic field. The second quantized Hamiltonian for this system in
terms of a fermionic field $\psi$ is given by
$$
H = \int dx \bigl[ {1 \over 2m} (D\psi(x))^{\dag} D\psi(x) + A_{0}(x) \psi
^{\dag} (x) \psi (x) \bigr] \eqno(42) $$
where $D$ is the covariant derivative $D= {\partial \over \partial x} -i A
(x)$.

It is known that excitations near the Fermi surface admit a relativistic field
theoretic description [18]. The corresponding relativistic Lagrangian is given
by
$$ L= \bar{\Psi} \gamma^{\mu} (i\partial _{\mu} - A_{\mu}) \Psi
\eqno(43) $$
where $\Psi = {\psi_{R} \choose \psi_{L}}$, $A^{\mu}=(A^{0},v_{F}A)$,
$x^{\mu}=(t, {x \over v_{F}})$, $\gamma^{\mu}=(\tau_{1},-i\tau_{2})$,
$\gamma^{5}=\gamma^{0}\gamma^{1}=\tau^{3}$, $g^{\mu\nu}=diag(1,-1)$ and
$\epsilon^{01}=1$.  $\psi_{L}$,
$\psi_{R}$ are the left and right moving components of the nonrelativistic
fermion field near the Fermi surface, $v_{F}$ is the Fermi velocity and
$\tau$'s
are the Pauli
matrices.
Equation (43) describes the fermion part of the Schwinger model Lagrangian. In
particular this model exhibits a chiral anomaly expressed by $$
\partial_{\mu}j^{\mu}_{5} = - \epsilon^{\mu\nu}\partial_{\mu}j_{\nu}={1 \over
{2\pi v_{F}}}\epsilon^{\mu\nu} F_{\mu\nu} \eqno(44) $$ where $ j^{\mu}_{5} =
\bar{\Psi}\gamma^{\mu}\gamma_{5}\Psi$ and   $j^{\mu} =
\bar{\Psi}\gamma^{\mu}\Psi$ [19].

We are now going to redescribe the
original system in terms of our collective field formalism. We want to find
out how the Lagrangian for the Schwinger model, eq.(43), and the corresponding
chiral anomaly, eq.(44), manifest themselves in the collective field
representation. The collective Hamiltonian is
$$\eqalignno{ H = & \int dx \Bigl[{1 \over 2m} \rho (x) \Bigl(\partial \pi (x)
 -A(x) \Bigr) ^{2}  + {1 \over 8m} {{(\partial
\rho (x))^ {2}} \over \rho (x)} + A_{0}(x) \rho (x) \cr
& + {\pi ^{2} \over 6m} \rho ^{3} (x)  \Bigr] -e(\int dx \rho (x)-N) + \cdots
 &(45) \cr} $$

We shall now employ a
semiclassical treatment by expanding around the time
independent classical solutions   $ \rho (x) = \rho _{0} (x) + \delta
\rho (x) $
where $\rho _{0} (x) $ satisfies
$$ \eqalignno{
& {1 \over 8} \Bigl[ \bigl({{\partial \rho _{0} (x)} \over \rho _{0} (x)}
\bigr) ^{2} -2{{\partial ^{2}\rho _{0} (x)} \over {\rho _{0}(x)}} \Bigr] +
{{\pi ^{2}} \over 2} \rho _{0}^{2}(x)-em=0 \cr
& \int \rho_{0}  (x) dx = N
&(46) \cr}$$  The most obvious solution is the constant solution $ \rho_{0} =
{N
\over L} ={p _{F} \over \pi}$.

The Hamiltonian for the excitations $\delta \rho$ is up to a constant
(assuming that the external electromagnetic field is weak and the fields are
slowly varying) $$ H= \int dx \bigl[ {{\rho _{0}} \over 2m} \partial \pi (x)
\partial \pi (x) -{{\rho _{0}} \over m}\partial \pi (x) A(x) +  A_{0}(x)
 \delta \rho (x)+ {{ \pi^{2}} \over 2m} \rho _{0}^{2} (\delta
\rho (x)) ^{2} \bigr] \eqno(47) $$  where $\delta \rho,~ \pi$ are
conjugate field variables with a subsidiary condition $\int \delta \rho (x)
dx =0$. This Hamiltonian can be derived from a local Lagrangian of the
form   $$ {\cal {L}}= {1 \over 8\pi} \partial _{\mu} \chi \partial ^{\mu}
\chi - {1 \over {2\pi \sqrt{v_{F}}}} \epsilon^{\mu\nu} A_{\mu} \partial_{\nu}
\chi \eqno(48) $$
where $\delta \rho ={1 \over 2} \sqrt{{\rho _{0}} \over {m \pi}}\partial \chi$.
The
relativistic notation we used has been indicated earlier. The above Lagrangian
is
essentially the bosonized form of its fermionic counterpart in eq.(43) [20].

The anomaly equation (44) is now expressed in terms of the equation of motion
for the field $\chi$. We can define the vector current $j_{\mu}$ as
$$j_{\mu} (x) \equiv -{\partial {\cal {L}} \over {\partial A^{\mu} (x)}} = {1
\over {2\pi \sqrt{v_{F}}}} \epsilon^{\mu \nu} \partial _{\nu} \chi(x) \eqno(49)
$$ Then using the equations of motion for $\chi$ we find that
$$ -\epsilon ^{\mu \nu} \partial_{\mu} j_{\nu} = {1 \over {2\pi
\sqrt{v_{F}}}}\partial_{\mu}
\partial^{\mu} \chi = {1 \over {2\pi v_{F}}} \epsilon^{\mu \nu} F_{\mu \nu}
\eqno(50) $$
which is the anomaly equation.

In the above analysis we considered only the constant solution of the static
classical equations (46) and the fluctuations around it. But eq. (46)
admits also a soliton solution, eq.(38). A semiclassical analysis around this
nontrivial configuration has to be done and appropriately interpreted in the
fermionic picture.

\vskip 1 cm
\noindent
{\bf {6. Discussion}}

In the previous section we talked about soliton solutions of the classical
equations of motion of the system of one-dimensional nonrelativistic fermions
in the collective field representation in the absence of an external potential.
It would be very
interesting to find if soliton-like solutions persist in the presence of an
external potential,
particularly an inverted harmonic oscillator potential, which is the case in
the double
scaling limit of the $D=1$ matrix model. The existence of new classical
solutions of the equations of motion of the collective Hamiltonian describing
the double scaling limit of the $D=1$ matrix model might make clearer the
connection between the field theory of the $D=1$ matrix model and the $D=1$
string theory.

\vfill \eject
\noindent
{\bf{7. Acknowledgements}}

I would like to thank Prof. B. Sakita for a fruitful collaboration on which the
results presented in this talk are based. I would also like to thank Antal
Jevicki for discussions. This research is supported by NSF grant PHY90-20495.

\vskip 1 cm
\noindent
{\bf{8. References}}

\item{1.} E. Br$\acute{\rm e}$zin, V.A. Kazakov and A.B. Zamolodchikov, {\it
Nucl.Phys.} {\bf B338} (1990) 673; G. Parisi, {\it Phys.Lett.} {\bf B238}
(1990)
209; D. Gross and N. Miljkovi$\acute{\rm c}$, {\it Phys.Lett.} {\bf B238}
(1990)
217; P. Ginsparg and J. Zinn-Zustin, {\it Phys.Lett.} {\bf B240} (1990) 333; J.
Polchinski, {\it Nucl.Phys.} { \bf B346} (1990) 253.
\item{2.} E. Br$\acute{\rm
e}$zin, C. Itzykson, G. Parisi and J.B. Zuber, { \it Comm.Math.Phys.} {\bf 59}
(1978) 35.
\item{3.} D. Karabali and B. Sakita, Preprint CCNY-HEP-91/2, to
appear in IJMP.
\item{4.} D. Karabali and B. Sakita, to appear in the proccedings
of the International Sakharov Conference, Moscow 1991.
\item{5.} D. Karabali and B. Sakita, in preparation.
\item{6.} R. Dashen and D. Sharp, {\it Phys. Rev.} {\bf 165} (1968) 1857; D.H.
Sharp, {\it Phys. Rev.} {\bf 165} (1968) 1867.
\item{7.} A perturbative calculation of the Jacobian, up to the first few
terms,
has been also given by A. Jevicki, {\it Nucl. Phys.} {\bf B146} (1978) 77.
\item{8.} A. Jevicki and B. Sakita, {\it Nucl. Phys.} {\bf B165} (1980) 511;
{\bf B185} (1981) 89.
\item{9.} A. Sengupta and S. Wadia, Tata Institute preprint, 90-33.
\item{10.} D.J. Gross and I. Klebanov, Princeton preprint, PUPT-1198.
\item{11.} V. Kazakov, preprint LPTENS 90/30.
\item{12.} G. Moore, preprint YCTP-P8-91, RU-01-12.
\item{13.} S.R. Das and A. Jevicki, {\it Mod. Phys. Lett.} {\bf A5} (1990)
1639.
\item{14.} J. Polchinski, University of Texas preprint UTTG-06-91; D. Minic, J.
Polchinski and Z. Yang, University of Texas preprint UTTG-16-91.
\item{15.} J. Avan and A. Jevicki, Brown preprint, Brown-HET-801.
\item{16.} A. Jevicki, Brown preprint, Brown-HET-807.
\item{17.} S.R. Das and A. Jevicki, unpublished, private communication with A.
Jevicki.
\item{18.} e.g. Z.B. Su and B. Sakita, {\it Phys. Rev.Lett.} {\bf 56} (1986)
780.
\item{19.} J. Schwinger, {\it Phys. Rev.} {\bf 128} (1962) 2425.
\item{20.} S. Coleman, {\it Phys.Rev.} {\bf D11} (1975) 2088; S. Mandelstam,
{\it Phys.Rev.} {\bf D11} (1975) 3026.

\bye